\documentstyle[12pt,aps,preprint]{revtex}
\begin{document}

\draft

\title{Universal Fluctuations in Spectra of Disordered Systems\\
at the Anderson Transition\thanks{to be appeared in Jpn. J. Appl. Phys.,
{\bf 34} (8A), 1995.}}

\author{Isa Kh.~{\sc Zharekeshev}~\cite{byline} and Bernhard~{\sc Kramer}}

\address{1 Institut f\"ur Theoretische Physik, Universit\"at Hamburg,
Jungiusstrasse 9, D--20355 Hamburg, Germany}

\date{21 February, 1995}

\maketitle

\begin{abstract}
Using the level--spacing distribution and the total probability
function of the numbers of levels in a given energy interval we analyze the
crossover of the level statistics between the delocalized and the localized
regimes.   By numerically calculating the electron spectra of systems of up
to $32^3$ lattice sites described by the Anderson Hamiltonian it is shown
that the distribution $P(s)$ of neighboring spacings is {\em scale--
independent} at the metal--insulator transition. For large spacings it has a
Poisson--like asymptotic form $P(s)\propto \exp (- A\,s/\Delta )$, where $A
\approx 1.9$. At the critical point we obtain a linear relationship between
the variance of the number of levels $\langle [\delta n(\varepsilon)]^2
\rangle $ and their average number $\langle n(\varepsilon)\rangle $ within
the interval $\varepsilon$. The constant of proportionality is less than
unity due to the repulsion of the levels. Both $P(s)$ and $\langle [\delta
n(\varepsilon)]^2 \rangle $ are determined by the probability density
$Q_{n}(\varepsilon)$ of having exactly $n$ levels in the energy interval
$\varepsilon $. The distribution $Q_{n}(\varepsilon)$ at the critical point
is found to be size--independent and to obey a Gaussian law near its maximum,
where $n\sim \langle n \rangle $.
\end{abstract}

\section{Introduction}
The statistical properties of the energy levels of disordered systems near
the metal-insulator transition
are presently attracting much attention.
Due to Anderson localization, the one-electron states of
systems with higher dimensionality ($d>2$) are known to experience
a crossover from delocalized to localized behavior.
This metal-insulator transition (MIT) is induced by increasing
the fluctuations of a random potential or by shifting the Fermi
energy through the mobility edge at fixed disorder.~\cite{KramerM94}
The statistics of the corresponding energy levels is very
sensitive to the MIT and depends dramatically on the strength of disorder.
In the insulating regime the levels are completely random,
and their distribution
obeys the Poissonian statistical laws of uncorrelated variables.
On the other side of the MIT, in the metallic regime,
correlations in the spectrum become
stronger due to quantum-mechanical level repulsion.
It was shown earlier~\cite{Efetov83,Altshuler86} that
the level statistics in a disordered metal is governed by the random-matrix
theory.~\cite{Wigner55,Dyson62,RMT}

By using finite--size scaling arguments
Shklovskii {\em et al}.~\cite{Shklovskii93}
suggested that exactly at the MIT
the distribution of spacings $P(s)$ between the neighboring energy
levels exhibits critical
behavior and
is a scale--invariant function which differs
considerably from both the Wigner surmise
$P_{W}(s)=(\pi s/2)\exp(- \pi s^{2}/4)$
and the Poisson law $P_{P}(s)=\exp(-s)$, corresponding
to the delocalized and the localized regime, respectively.
Here $s$ is measured in units of the mean level spacing $\Delta$.
Later the universal properties of $P(s)$
at the mobility edge
were studied
in some detail by using approximate analytical~\cite{Aronov94,Kravtsov94}
and numerical methods.~\cite{Evangelou94,Hofstetter94,ZharekeshevK94}

In this paper we present results of elaborate
numerical investigations of the
statistics of fluctuations in the energy spectra of three--dimensional (3D)
disordered lattices near the MIT.
Using the Anderson model we calculate very accurately
not only the critical distribution
$P(s)$ of neighboring spacings, but also
the total probability density $Q_{n}(\varepsilon)$
of having $n$  levels within an
energy interval $\varepsilon$. The latter contains
more complete information about spectral correlations of higher orders.
Both distributions exhibit a universal character at the critical disorder
and become independent of the system size.
We find that the variance of the number of the levels
in an interval $\varepsilon$ as a function of mean number of levels
is also scale--invariant, and obeys  a linear relation
$\langle \delta n^2 \rangle = \kappa  \langle n \rangle$ with
$\kappa \approx 0.27$, when $\langle n \rangle \gg 1$.

\section{Nearest-Neighbor Level Spacing Distribution}
In order to calculate the spectrum we use the Hamiltonian corresponding
to a tight-binding
model with diagonal disorder,
$H=\sum_{n}\varepsilon_{n}^{} a_{n}^{\dag} a_{n}^{} +
          \sum_{n\neq m} (a_{n}^{\dag} a_{m}^{} + c.c.)$,
where $a_{n}^{\dag}$ ($a_{n}^{}$) is the creation (annihilation)
operator of an electron at a site $n$ in a
simple cubic lattice.
The random site energy $\varepsilon_{n}$
is distributed uniformly between  $-W/2$ and $W/2$.
The second sum involves only pairs of nearest sites \{$n,m$\}.
The MIT in the center of the band corresponds to
$W_{c}\approx 16.5$.~\cite{Kramer83}

The energy levels of cubes
of various linear
sizes from $L=5$ to 32 with periodic boundary conditions
were calculated by direct diagonalization
of the Hamiltonian
using an optimized Lanczos algorithm.~\cite{ZharekeshevK94}
%
%
\marginpar{\fbox{Fig. 1}}
Figure~\ref{fig1}\, displays the histogram of $P(s)$ calculated
at the disorder corresponding to the critical point, $W_{c}$.
The number of realizations was chosen so that the numbers of levels
from central half part of the band were approximately
$1.2\,10^{7}$, $1.2\,10^{6}$ and $2\,10^{5}$ for $L=5$, 12 and 32,
respectively.
All of the obtained data, regardless of $L$, lie on
the same curve,
the critical $P(s)$, which is
considerably different from both $P_{W}(s)$ and $P_{P}(s)$.
However, for disorder $W\neq W_{c}$ the distribution
$P(s)$ depends on the system size and exhibits scaling behavior
near the MIT.~\cite{Shklovskii93}
Indeed, for decreasing $W$ the
distribution of spacings approaches the Wigner surmise,
while for increasing $W$ it scales towards a Poissonian law.
This crossover is apparently
accompanied by the transition from the delocalized to
the localized regime.
The finite--size scaling properties of $P(s)$ allow
determination of the critical exponent $\nu \approx 1.45$
and the disorder dependence of the scaling parameter,
the correlation length $\xi(W)$.~\cite{Hofstetter94,ZharekeshevK94}
We have also verified that the form of $P(s)$ exactly at the MIT
does not depend on the width of
the energy interval as long as the levels in this interval belong
to the critical region, i.e. satisfy the condition
$L<\xi=(|\varepsilon - \varepsilon_{c}|/\varepsilon_{c})^{-\nu}$, where
$\varepsilon_{c}$ is the mobility edge.

Using a large number of realizations of the randomness
we were able to analyze
the asymptotic behavior of the critical $P(s)$
as shown in the inset of Fig.~\ref{fig1}.
For spacings $s>2$ the exponential tail of $P(s)$ is intermediate
between the Gaussian and the Poissonian. It is well described by
the relation $P(s) = \exp(-A s)$ with a
numerical coefficient $A\approx 1.9$.
One observes that this asymptotic decay deviates from the power-law
$\ln P(s) \propto s^{1+1/3 \nu}$
which was recently
found~\cite{Aronov94} by using the effective ``plasma model''
as defined by Dyson.~\cite{Dyson62}

\section{Probability of $n$ Levels in a Given Interval}
In order to analyze the correlations between several consecutive eigenvalues,
one can
study the probability that an
energy interval of width $\varepsilon$ centered  at a
randomly chosen energy contains exactly $n$ levels.
The distribution of this joint probability $Q_{n}(\varepsilon)$ is related
to the $n$-level correlation function
$R_{n}(\varepsilon_{1},\varepsilon_{2},...,\varepsilon_{n})$.
It provides a more complete description of
the level statistics than the two-point
correlation function,~\cite{RMT}
$R_{2}(\varepsilon_{1}-\varepsilon_{2})$, which can
be expressed in terms of the $n$-level spacing distribution $p(n,s)$,
$R_{2}(s)=\sum_{n=0}^{\infty} p(n,s)$.
On the other hand, the probability of finding no levels ($n=0$)
inside the interval defines the nearest-neighbor spacing distribution,
$P(s) = d^{2} Q_{0}(s)/ds^{2}$.

The results of the random--matrix theory for $Q_{n}(\varepsilon)$,
which correspond to the metallic case,
have been calculated numerically,~\cite{RMT} and were later
expressed in an explicit analytical form~\cite{ShklovskiiF94} for the three
universality classes of random Hamiltonians:
orthogonal, unitary and symplectic.
In the limit $\delta n \equiv |n-\varepsilon|\ll \varepsilon$
the distribution for the orthogonal ensemble is approximately Gaussian,
\begin{equation}
\ln Q_{n}(\varepsilon) \propto -\frac{\pi^{2}}{4}\,
\frac{\delta n^{2}}{\ln(8\varepsilon /|\delta n|)+B},
\label{top}
\end{equation}
where $\varepsilon$ is measured in $\Delta$
and $B$ depends weakly on $\delta n$.

In the insulating regime the sequence of the levels is completely random
due to the localization of the states, and
therefore we have the usual Poissonian process,
\begin{equation}
Q_{n}(\varepsilon) = \varepsilon^{n} \exp(-\varepsilon)/n!.
\label{process}
\end{equation}
For example, if $n=0$ the Wigner surmise and the Poisson law give
\begin{equation}
Q_{0}(\varepsilon) =
1- \frac{2}{\sqrt{\pi}}\,{\rm Err}(\frac{\sqrt{\pi}}{2} \varepsilon), \qquad
Q_{0}(\varepsilon) = \exp(-\varepsilon)
\label{Q0}
\end{equation}
for the metallic and insulating regime, respectively.
Err$(x)$ is the error function.

It is of great interest to study how $Q_{n}(\varepsilon)$ changes
from~(\ref{top}) to~(\ref{process}) with increasing disorder $W$,
and to calculate its shape at the MIT.
Our results suggest that the function $Q_{n}(\varepsilon)$ shows
critical behavior near $W_{c}$.
In Fig.~\ref{fig2} we show $Q_{n}(\varepsilon)$
corresponding to the critical point.
\marginpar{\fbox{Fig. 2}}
Calculations for different system sizes at $W_{c}$
yield almost the same set of
distributions independently of the number of levels $n=0, 1, 2,...$.
The width of $Q_{n}(\varepsilon)$ at the MIT is larger than in the
metallic regime, since the level repulsion becomes less important.
We note that for small fluctuations ($\delta n \ll \varepsilon$)
$\ln Q_{n}(\varepsilon)$ is quadratic with respect to both $\varepsilon$
and $n$, similarly to ~(\ref{top}).

Let us consider the quantity
$I_{n} = \int_{o}^{\infty} Q_{n}(\varepsilon) d\varepsilon$.
For the Gaussian orthogonal ensemble
it is known from previous numerical simulations~\cite{RMT}
that $I_{0} \approx 0.643$, $I_{1} \approx 0.922$,
and  $I_{n}$ converges to unity in the limit
$n\to \infty$.
For the Poissonian process~(\ref{process})
it is easy to see  that $I_{n}=1$ for all $n$.
At the MIT, when $W=16.5$, we obtained the following values of
the integral:
$I_{0}^{c} \approx 0.714$,
$I_{1}^{c} \approx 0.960$, $I_{2}^{c} \approx 0.994$,
$I_{3}^{c} \approx 0.998$, and also $\lim_{n\to\infty} I_{n}^{c}=1$.
The latter set of constants is $L$-independent and characterizes
the critical level statistics.
We emphasize that $I_{n}$ changes with $W$ around the
critical disorder $W_{c}$ according to a one--parameter scaling law,
$I_{n}(W,L) = I_{n}^{c} + f(L/\xi(W))$, where the function $f(x)$ can
be linearized in the vicinity of the critical point,
$f(L,W) \propto (W-W_{c}) L^{1/\nu}$. This allows determination of the
critical exponent $\nu$ of the correlation length. Note that in searching
for $\nu$ it is not necessary to choose
a certain spacing $s_{o}$ as was done earlier when using
$P(s)$.~\cite{Hofstetter94,ZharekeshevK94}

\section{Variance of the Level Number}

The width of the distribution $Q_{n}(\varepsilon)$ describes the
rigidity of the spectrum, which is defined by the variance
of the level number,
$\langle [\delta n(\varepsilon)]^2 \rangle =
\sum_{n=0}^{\infty}(n-\langle n
\rangle)^{2}Q_{n}(\varepsilon)$,
where $\langle n \rangle = \varepsilon$ is the average number of
levels in a given  interval $\varepsilon$.
One can ask how dependence of the variance
$\langle [\delta n(\varepsilon)]^2 \rangle$
on the average $\langle n(\varepsilon)\rangle $ changes
when the delocalized
states transform into the localized ones.

It follows from the random--matrix theory~\cite{Altshuler86,RMT} that
in the metallic regime the variance
is given by the Dyson formula,
\begin{equation}
\langle \delta^2 n \rangle_{\rm M}
= \frac{2}{\pi^2}(\ln \langle n \rangle +  C),
\qquad 1 \ll \langle n \rangle \ll \hbar D/L^{2},
\label{Dyson}
\end{equation}
where $C \approx 2.18$ and $D$ is the diffusion constant.
By increasing the disorder $W$
the fluctuations of $\delta n$ are increased due to a
weakening of level correlations.
In the strongly localized regime the levels are not correlated, hence
$\langle \delta^2 n \rangle_{\rm I} = \langle n \rangle$.
This is much larger than $\langle \delta^2 n \rangle_{\rm M}$.
Exactly at the MIT, the variance depends linearly on
the average level number,~\cite{Altshuler88}
$\langle \delta^2 n \rangle_{c} = \kappa \langle n \rangle$, as in the
insulating limit. However
the factor $\kappa$ is less than unity.

We calculated the dependence of the ratio
$\langle \delta n^2 \rangle/\langle n \rangle$
on the average number of levels within a given interval $\varepsilon$
for different
lattice sizes $L$ at $W_{c}=16.5$
as shown in Fig.~\ref{fig3}.
\marginpar{\fbox{Fig. 3}}
Obviously the ratio is not sensitive to a change of $L$.
Such behavior results from the above--mentioned
{\em universality} of the probability density $Q_{n}(\varepsilon)$.
One observes  that
the amplitude of the relative fluctuations
of the number of levels
$\langle \delta^2 n \rangle/\langle n \rangle$
decreases very weakly with the
energy, approaching the constant limit
$\kappa \approx 0.27$.
Thus, our results are consistent with the suggestion of
proportionality between the variance and
the average level number at large $\langle n \rangle$,~\cite{Altshuler88}
but deviate from the recently proposed power law,~\cite{Kravtsov94}
$\langle \delta^2 n \rangle/\langle n \rangle
\propto \langle n \rangle^{-1/3\nu}$.
However, this power law can be considered in our calculations as the
next higher-order correction to the leading constant term.

It was earlier conjectured~\cite{Altshuler88}  that
$Q_{n}(\varepsilon)$ at the MIT is
normally distributed
{$\propto\exp[-(\varepsilon-n)^{2}/2\langle \delta^2 n \rangle_{c}]$}
for fixed $\varepsilon \gg n$.
After substituting $k\varepsilon$ for the variance
and doubly differentiating, for $n=0$
one recovers the Poissonian asymptotic form of the critical
$P(\varepsilon) \propto \exp(-\varepsilon/2\kappa)$.

\section{Conclusions}
We have discussed the statistics
of the spectral fluctuations
in the 3D systems at the disorder--induced metal--insulator transition.
By directly diagonalizing the Anderson Hamiltonian
the nearest-neighbor level spacing
distribution $P(s)$ and the probability $Q_{n}(\varepsilon)$ of finding
exactly $n$ eigenvalues in an energy interval of width $\varepsilon$
were numerically calculated.
We confirmed that the critical $P(s)$ is $L$-independent and has
a Poissonian like asymptotic form at large $s$, but with stronger
decay than that in the insulating regime.
One expects that the critical $Q_{n}(\varepsilon)$ which
describes the statistical fluctuations  in the discrete spectrum
exhibits also complete {\em scale--invariance} in analogy with the
universality of the critical $P(s)$.
Our results show that the distributions $Q_{n}(\varepsilon)$ at the MIT
are well described by Gaussians both for small fluctuations
$\delta n \ll \varepsilon$ and also when $n\ll \varepsilon$, as
in the metallic limit.
We  also analyzed
the variance of the number of states $\langle \delta^2 n \rangle$
in a given interval as a function of  mean level number $\langle n \rangle$.
At the critical point this function was found
to be independent of the system size and to have a leading linear term,
$\langle \delta^2 n\rangle_{c} = 0.27 \langle n \rangle$ for
$n\gg 1$.
To summarize, our results indicate that the statistics of the energy
levels near
the Anderson transition constitute an infinite set of scaling variables
$I_{n}$ which can be used to characterize the mobility edge
by one--parameter scaling laws.

\acknowledgements
The authors thank B.~I.~Shklovskii for discussions. Financial support
from DAAD during the stay of I.~Kh.~Zh. at the University of
Hamburg is gratefully acknowledged.

\begin{figure}
\caption[]{Level spacing distribution $P(s)$ at the MIT for various system
sizes $L$. Curves: Wigner and Poisson distributions.
Inset shows the tail of $P(s)$. Straight line is best fit.
Dotted line: the power-law asymptotic~\cite{Aronov94}
$\ln P(s) \propto s^{1+1/3 \nu}$.}
\label{fig1}
\end{figure}

\begin{figure}
\caption[]{
Critical probability distribution $Q_{n}(\varepsilon)$
for different $n$.}
\label{fig2}
\end{figure}

\begin{figure}
\caption[]{
Variance of the level number $\langle \delta n^2 \rangle $
as a function of the average $\langle n \rangle $ for
various $L$. Solid line: Dyson result~(\ref{Dyson}).}
\label{fig3}
\end{figure}


\begin{thebibliography}{9}

\bibitem[\S]{byline}
Permanent address: Department
of Physics, Kazakh State University,
Almaty, Kazakhstan. E-mail address: isa@physnet.uni-hamburg.de


\bibitem{KramerM94}
For review, see
B. Kramer and A. MacKinnon: Rep. Prog. Phys. {\bf 56} (1994) 1469
and references therein.

\bibitem{Efetov83}
K.~B.~Efetov: Adv. Phys. {\bf 32} (1983) 53.

\bibitem{Altshuler86}
B.~L.~Altshuler and B.~I.~Shklovskii:
Sov. Phys. JETP {\bf 64} (1986) 127.

\bibitem{Wigner55}
E.~P.~Wigner: Ann. Math.\ {\bf 62} (1955) 548.

\bibitem{Dyson62}
F.~J.~Dyson: J. Math. Phys. {\bf 3} (1962) 140.

\bibitem{RMT}
M.~L.~Mehta: {\em Random Matrices} (Academic Press, Boston, 1991).

\bibitem{Shklovskii93}
B.~I.~{Shklovskii, B.~Shapiro}, B.~R.~Sears, P.~Lambrianides
and H.~B.~Shore: Phys. Rev. B {\bf 47} (1993) 11487.

\bibitem{Aronov94}
A.~G.~Aronov, V.~E.~Kravtsov and I.~V.~Lerner:
JETP Lett. {\bf 59} (1994) 40.

\bibitem{Kravtsov94}
V.~E.~Kravtsov, I.~V.~Lerner, B.~L.~Altshuler and A.~G.~Aronov:
Phys. Rev. Lett. {\bf 72} (1994) 888.

\bibitem{Evangelou94}
S.~N.~Evangelou: Phys. Rev. B {\bf 49} (1994) 16805.

\bibitem{Hofstetter94}
E.~Hofstetter and M.~Schreiber: Phys. Rev. B {\bf 49} (1994) 14726.

\bibitem{ZharekeshevK94}
I.~{Kh.~Zharekeshev} and B.~Kramer: to be published in Phys. Rev. B (1995).

\bibitem{Kramer83}
A.~{Mac\,Kinnon} and B.~Kramer: Z. Phys. B {\bf 53} (1983) 1.

\bibitem{ShklovskiiF94}
M.~M.~Fogler and B.~I.~Shklovskii: to be published in Phys. Rev. Lett. (1995).

\bibitem{Altshuler88}
B.~L.~Altshuler, {I.~Kh.~Zharekeshev}, S.~A.~Kotochigova and B.~I.
Shklovskii: Sov. Phys. JETP {\bf 67} (1988) 625.

\end{thebibliography}
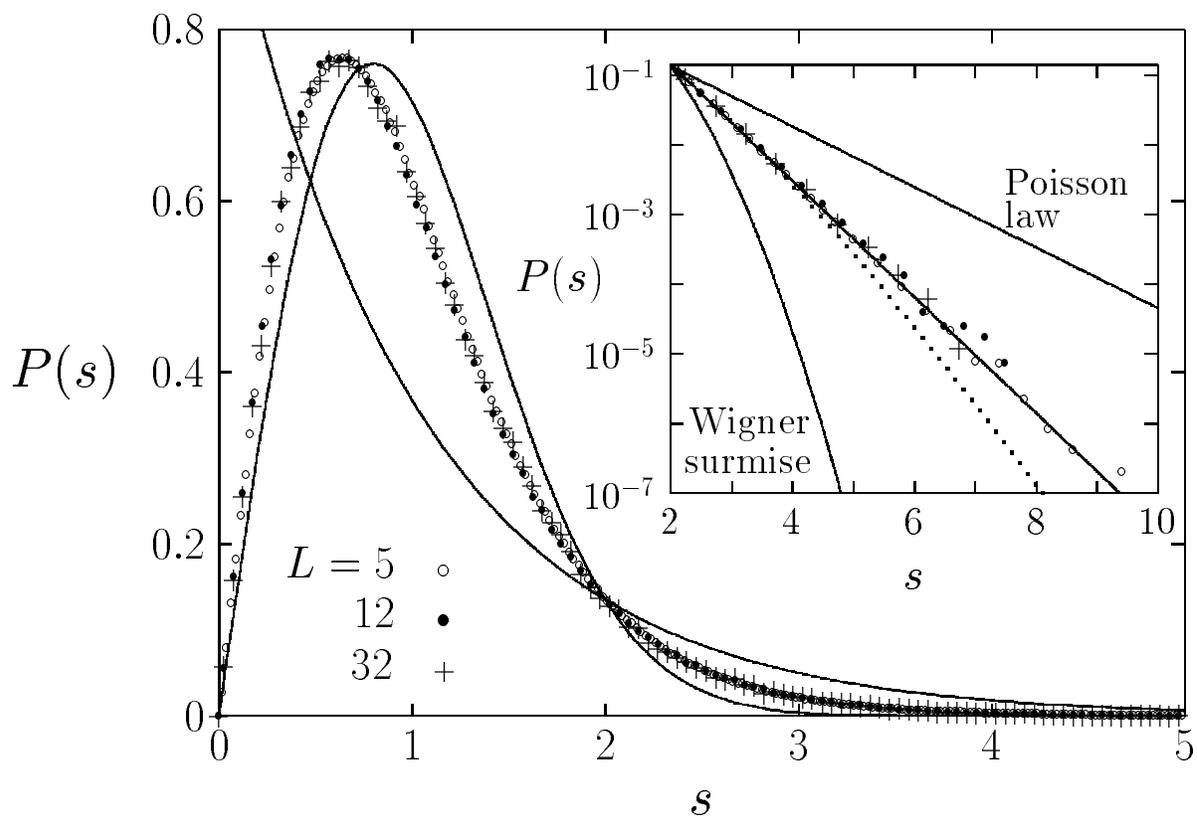
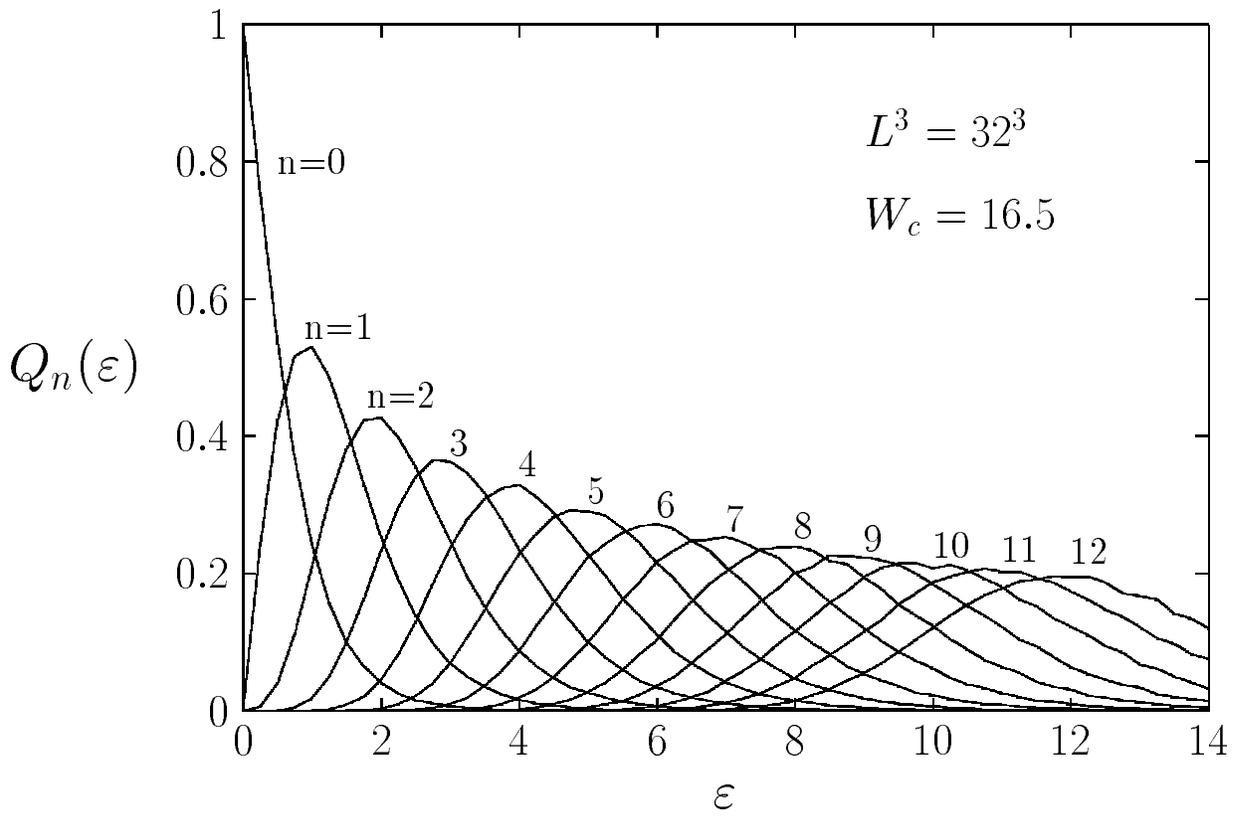
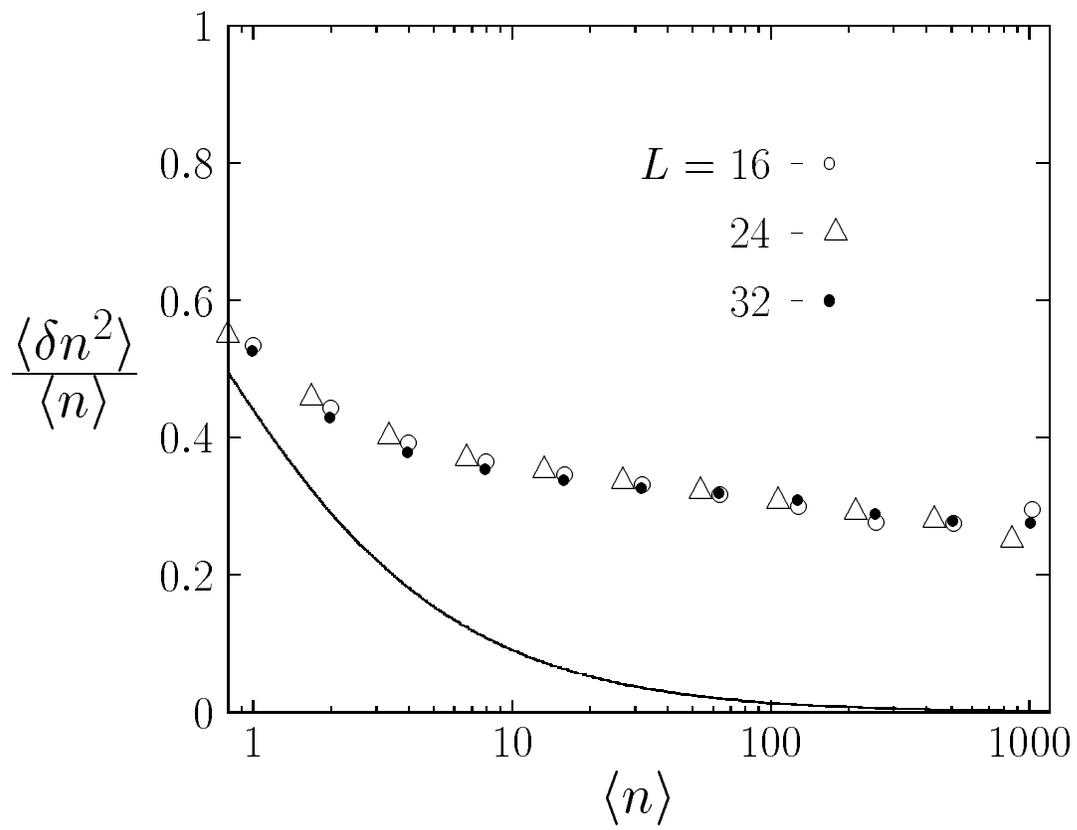
\end{document}